\documentclass{article} 
\usepackage{epsfig} 
\usepackage{bbm} 
\usepackage{latexsym}

\title{A Recursive Definition of the \\ Holographic Standard
  Signature} \author{William F. Bradley} 

\newtheorem{lemma}{Lemma}
\newtheorem{theorem}{Theorem}
\newtheorem{corollary}{Corollary}

\begin{document}
\bibliographystyle{plain}
\maketitle

\newcommand{\fa}{\ensuremath{\tilde{f}}}
\newcommand{\FF}{\ensuremath{\mathbbm{F}}}
\newcommand{\RR}{\ensuremath{\mathbbm{R}}}
\newcommand{\CC}{\ensuremath{\mathbbm{C}}}
\newcommand{\ua}{\ensuremath{\underline{\alpha}}}
\newcommand{\up}{\ensuremath{\underline{p}}}
\newcommand{\uf}{\ensuremath{\underline{f}}}
\newcommand{\ug}{\ensuremath{\underline{g}}}
\newcommand{\uh}{\ensuremath{\underline{h}}}
\newcommand{\ux}{\ensuremath{\underline{x}}}
\newcommand{\SSS}{\ensuremath{\mathbbm{S}}}
\newcommand{\specific}{\ensuremath{\hat{x}}}
\newcommand{\uspec}{\ensuremath{\underline{\specific}}}
\newcommand{\hA}{\ensuremath{\hat{A}}}
\newcommand{\hB}{\ensuremath{\hat{B}}}
\newcommand{\pf}{\mbox{Pf}}
\newcommand{\pM}{\mbox{PerfMatch}}

\newcommand{\eop}{~\ensuremath{\mathbbm{\Box}}}
\newcommand{\proof}{\textsc{Proof: }}

\abstract{ We provide a recursive description of the signatures
  realizable on the standard basis by a holographic algorithm.  The
  description allows us to prove tight bounds on the size of planar
  matchgates and efficiently test for standard signatures.  Over
  finite fields, it allows us to count the number of $n$-bit standard
  signatures and calculate their expected sparsity.}

\tableofcontents

\pagebreak
\section{Introduction}
Holographic algorithms have been a subject of much interest in the
mathematical community since Leslie Valiant conceived of them in
2002 (see~\cite{valiant1}).  These algorithms can calculate certain
exponential sums in polynomial time, skating dangerously close to
$\#P$ problems.

This paper will examine one small aspect of holographic algorithms;
our narrow focus will allow us to avoid some of the details and much
of the terminology surrounding the subject.  However, to provide a
little context for the reader unfamiliar with holographic computing,
we will give an extremely rough sketch of the subject in the next
paragraph.  More precise and complete introductions can be found
in~\cite{valiant3} or~\cite{cai1}.

We can think of holographic computing as follows.  Fix a field \FF.
Imagine that we build a circuit board out of special circuit
components.  Each component has a certain number of wires which we can
attach to other components.  We attach the wires so that none of them
cross on the circuit board.  (In other words, if we treat the circuit
components as nodes and the wires as edges, we form a planar graph.)
Each wire can take on only two values, either zero or one.  If we
specify the values of the wires attached to a component, it produces
an output value lying in \FF.  (If a component has $n$ wires, this
function from $\{0,1\}^n$ to $\FF$ is the ``signature'' referred
to in the title; we would call it an ``$n$-bit standard signature''.)
If we set all the wires on the entire circuit board, we define the
entire circuit board as producing the product of the outputs of the
individual circuit components.  A holographic algorithm lets us
compute the sum of these products over all (exponentially many) wire
settings in polynomial time.

If the signatures could be chosen freely, it would follow that
$P=\#P$.  Sadly, if not surprisingly, we lack this freedom: only some
functions are hospitable to holographic manipulations.  These special
functions are said to be \emph{realizable on the standard basis}, or
are simply called the \emph{standard signatures}.  It is possible to
change our computational basis, which produces new sets of signatures.
Much of the power of holographic algorithms arise from these changes
of basis; however, this paper focuses only on the simpler case of the
standard basis.

So, which functions are standard signatures?  Three equivalent
definitions are frequently used.  Standard signatures were originally
defined in terms of sums of weighted matchings on planar graphs by
Valiant in~\cite{valiant2}.  However, Cai and Choudhary established an
equivalence between standard signatures and the Pfaffians of certain
matrices in~\cite{cai3} and~\cite{cai4}, providing a second
definition.  One consequence of their result is a description of the
standard signatures as an algebraic variety: a function is a standard
signature if and only if a certain set of quadratic equations evaluate
to zero.  This provides a third definition of a standard signature.

Although the reader may think that three definitions is more than
enough, we offer a fourth one.  Our ``new'' definition is really a
consequence of the Pfaffian definition, but it seems to highlight
different properties than the other definitions.  Our definition is
recursive, i.e.\ we define $n$-bit standard signatures in terms of
$(n-1)$-bit standard signatures.  Here are some of the conclusions we
draw:
\begin{itemize}
\item
If we are operating over a finite field, we can count the exact number
of $n$-bit standard signatures.  We can also calculate the asymptotics
for large $n$.  Over $\FF_2$ and $\FF_3$, the number of odd parity
standard signatures coincides with the number of $n$-dimensional
self-dual codes.  (See Subsections~\ref{subsec:count},
\ref{subsec:asymptotics}, and~\ref{subsec:self.dual}, respectively.)
\item
It is known that any $n$-bit standard signature can be represented by
a planar matchgate with at most $O(n^2)$ nodes.  We construct a
matching lower bound showing that there exist standard signatures that
require at least $\Omega(n^2)$ nodes to encode as a planar
matchgate. (See Subsection~\ref{subsec:matchgate.size}.)
\item
Suppose we are given an $n$-bit function and we would like to
determine if it is a standard signature.  The naive approach takes
$O(n2^{2n})$ steps; using recursion and some structural properties, we
can improve this bound to $O(n2^n)$ steps.  (See
Subsection~\ref{subsec:complexity}.)
\item
Suppose we are working over a finite field and we select an $n$-bit
standard signature $f$ uniformly at random.  We can calculate the
expected sparsity of $f$, i.e.\ $\Pr[f(x)\neq 0]$.  (See
Subsection~\ref{subsec:sparse}.)
\end{itemize}

The paper is structured in two halves.  In the first half,
Section~\ref{sec:def}, we present the four different definitions of a
standard signature and a few lemmas.  In the second half,
Section~\ref{sec:hints}, we illustrate various corollaries of the
recursive definition.  Subsections~\ref{subsec:expressiveness}
and~\ref{subsec:self.dual} are more speculative in nature.  We also
include two appendices: Appendix~\ref{app:6bit} lists the general form
for a normalized 6-bit standard signature, and
Appendix~\ref{app:matchgate} illustrates one method of building
recursion into planar matchgates.

\section{Definitions} \label{sec:def}
Let $V=\{0,1\}$ be the field with 2 elements.  We will be considering
functions from $V^n\rightarrow \FF$, where $\FF$ is an arbitary field.
We refer to these as $n$-bit functions.  (Other authors would call
them $n$-arity functions.)  Given $x\in V^n$, we often expand it in bits
as $x=x_1\cdots x_n$.

To keep our notation saner, if $\alpha$ is a bit string and we remove
a bit from it, we will write $\ua$.  In a similar vein, given a
function $f:V^n\rightarrow \FF$, we can fix the last bit and define a
new function $\uf_0:V^{n-1}\rightarrow \FF$ as
\[\uf_0(x_1\cdots x_{n-1})=f(x_1\cdots x_{n-1}0)\]
and
\[\uf_1(x_1\cdots x_{n-1})=f(x_1\cdots x_{n-1}1)\]

Let $e_i\in V^n$ be the string all of whose bits equal zero except for
the $i$-th bit.  Also, for any two $n$-bit strings $x$ and $y$, let
$x+y$ represent the bitwise XOR of the two strings.

Given $x=x_1x_2\cdots x_n\in V^n$, let $|x|$ be the \emph{Hamming weight} of
$x$, i.e.\ 
\[|x|=x_1 + x_2 + \cdots + x_n\]
We define the \emph{partial Hamming weight} as follows:
\[|x|_j^k = \sum_{i=j}^kx_i\]
Note that $|x|_1^n=|x|$.  If $k<j$, define $|x|_j^k=0$.

If $f(x)=0$ for all $x$, we call $f$ the \emph{constant zero
  function}, and write $f\equiv 0$.  We refer to other functions as
\emph{non-zero functions}, or write $f\not\equiv 0$.

We can interpret the input either as an $n$-bit string, or as (the
binary representation of) an integer in the range $[0,2^n-1]$.  Using
the integer representation, we can specify a function
$f:V^n\rightarrow \FF$ by listing its outputs (i.e.\ its ``truth
table'').  That is, $f$ is fully determined by the ordered list
\[(f(0), f(1), f(2),...,f(2^n-1))\in \FF^{2^n}\]
Viewed as elements of $\FF^{2^n}$, functions form a vector space over
$\FF$: we can add together two functions, and we can multiply them by
scalars in $\FF$.

We say that a function $f:V^n\rightarrow \FF$ has \emph{even parity}
if all odd weight codewords are sent to zero, that is
\[\mbox{if }|x|=1 \bmod 2 \mbox{ then } f(x)=0\]
If $f$ has even parity and is not the constant zero function, then we
say that $f$ is \emph{strictly even parity}.  We can define
\emph{(strictly) odd parity} functions in the same way.  Note that the
constant zero function is the unique $n$-bit function that has both
even and odd parity.

\subsection{Standard Signatures via Planar Matchgates} \label{subsec:graph}
In this section, we will define a class of functions, the
\emph{standard signatures}, in terms of certain graphs and
perfect matchings.

A \emph{planar matchgate} over $\FF$ is a planar embedding of a planar
graph $G$ with weighted edges $w_{i,j}\in\FF$, along with a set of
special ``input/output'' nodes $v_1,...,v_n$ on the outer face of the
graph.\footnote{In a more typical definition, as in~\cite{valiant2},
  the input/output nodes are divided into distinct sets of ``input''
  and ``output'' nodes.  However, as long as we restrict our attention
  to the standard basis, that distinction is irrelevant, so we skip it
  for this paper.}  We label the index of each $v_i$ consecutively;
that is, if we start at node $v_i$, and proceed in an anti-clockwise
direction around the outer face, the next input/output node we
encounter is $v_{i+1}$.

We give an example below where $\FF=\RR$.  The small numbers are the
edge weights, the large numbers are the labels of the input/output
nodes. Two of the outer nodes are not input/output nodes (and thus are
not labelled):
\begin{center}
\epsfig{file=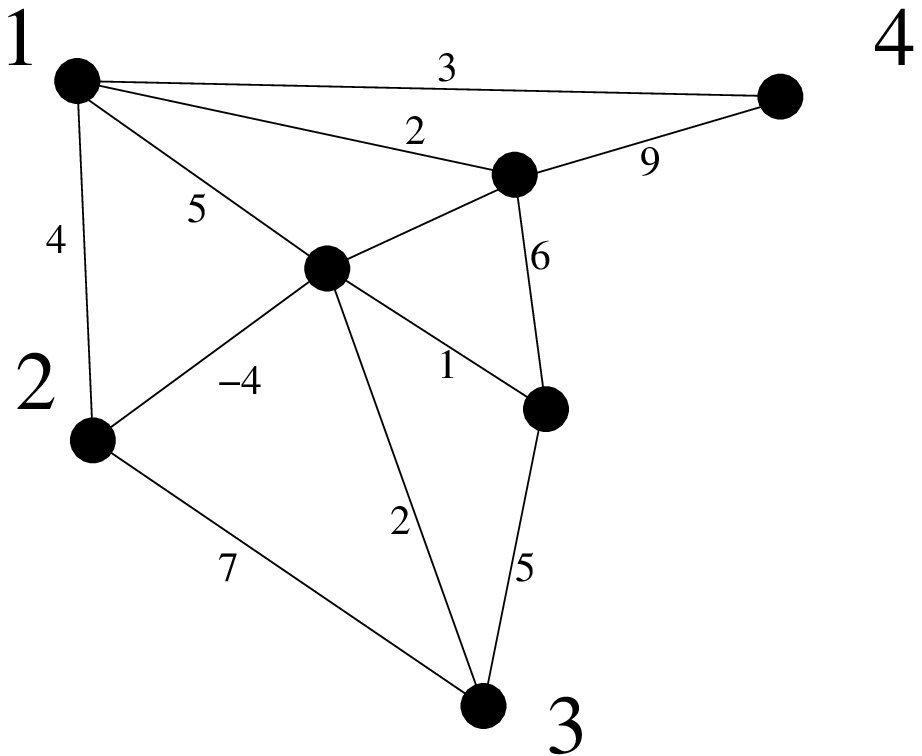, height=1.5in}
\end{center}

A \emph{perfect matching} is a collection of edges $E$ such that every
node is adjacent to exactly one edge in $E$.  The \emph{weight} of a
particular perfect matching is the product of the weights of the edges
in $E$.  Following Valiant, we will define $\pM(G)$ to be the
sum of the weight of every perfect matching in $G$ (or zero if there
are none.)  In other words,
\[\pM(G)=\sum_{E}\prod_{(i,j)\in E}w_{i,j}\]

Next, specify a vector $x\in \{0,1\}^n$.  If the $i$-th bit of $x$ is
a one, then suppose we remove node $v_i$ and all of its adjacent edges
from $G$.  This produces some subgraph, which we will call $G_x$.  We
can now define a function $f:\{0,1\}^n\rightarrow \FF$ by
\[f(x) = \pM(G_x)\]
The set of functions that can be described in this fashion (for some
$G$) form the \emph{$n$-bit standard signatures over $\FF$}.

Given a weighted planar graph $G'$, it is possible to calculate
$\pM(G')$ in time polynomial in the number of nodes using an object
called a Pfaffian.  This result was proved by Fisher, Kasteleyn and
Temperley in 1961 (see~\cite{kasteleyn} for a survey); this is
sometimes called the FKT Theorem.  We will examine Pfaffians in
greater detail in Subsection~\ref{subsec:pfaff}.

We will need some notation to describe various sets of standard
signatures.  First, let $A_n$ be the set of $n$-bit standard
signatures.  (The set $A_n$ depends on $\FF$ of course, but we will
treat $\FF$ as constant, so we will suppress the extra notation.)  
We can partition $A_n$ into three disjoint subsets, based on
the parity of the function:
\[A_n=A_n^{odd} \cup A_n^{even} \cup A_n^{0}\]
where $A_n^{odd}$ consists of the strictly odd parity standard signatures,
$A_n^{even}$ consists of the strictly even parity standard signatures, and
$A_n^0$ is a one-element set consisting of the constant zero function.

We will find it useful to normalize the standard signatures.  Let us
define a \emph{normalized standard signature} as a standard signature
$f$ where $f(0\cdots0)=1$.  We let $B_n$ be the set of normalized
standard signatures.  Note that all the elements of $B_n$ are strictly
even parity.\footnote{It might seem more natural to define a function
  as normalized if $f(1\cdots 1)=1$.  However, the parity would change
  as a function of $n$; our definition makes the parity of $B_n$ even
  for all $n$.}

\subsection{Basic Lemmas}

Before continuing with our definitions, we mention
a few lemmas that we will find useful later.

\begin{lemma} \label{lem:zero}
If $f\equiv 0$ then $f$ is a standard signature.
\end{lemma}

\proof Given any $n$-bit planar matchgate, we can add two more nodes
and an edge between them of weight 0; the resulting standard signature
is identically zero.\eop

For $n=1$, we can write down $A_1^{odd}$ and $A_1^{even}$ explicitly.
We will state it as a lemma for future reference.
\begin{lemma} \label{lem:a1}
We can characterize the 1-bit standard signatures over any field $\FF$:
\begin{eqnarray*}
A_1^{odd}=\{f\in V\rightarrow \FF \, | \, f(0)=0 \mbox{
  and } f(1)\neq 0\}\\
A_1^{even}=\{f\in V\rightarrow \FF \, | \, f(1)=0 \mbox{
  and } f(0)\neq 0\}\\
A_1^{0}=\{f\in V\rightarrow \FF \, | \, f(0)=0 \mbox{
  and } f(1)= 0\}
\end{eqnarray*}
\end{lemma}

\begin{lemma}\label{lemma:involute}
By flipping a fixed input bit, we can construct a bijection between
strictly even and strictly odd standard signatures.
\end{lemma}

\proof Suppose we have a planar matchgate and node $v$ is labelled as
the $i$-th input/output node.  Suppose we add a new node $v'$, an edge
between $v$ and $v'$, and we relabel node $v'$ as the $i$-th
input/output node.  If $f(x)$ is the standard signature of the
original planar matchgate, then $f(x+e_i)$ is the standard signature
of the new planar matchgate.  Note that $f(x)$ and $f(x+e_i)$ have
opposite parities.  Since this operation (flipping the $i$-th bit) is
invertible, we have established our bijection.\eop

Next, we let us examine normalized functions more carefully.
Normalization preserves the quality of being a standard signature:
\begin{lemma} \label{lem:normalize}
Suppose that $f:V^n\rightarrow \FF$ and there exists $\specific$ such
that $f(\specific)=\beta\neq 0$.  Let 
\[g(x)=\beta^{-1}f(x+\specific)\]
(Note that $g(0\cdots 0)=1$.)  Then $f$ is a standard signature if and
only if $g$ is a standard signature.
\end{lemma}

\proof Suppose $f$ is a standard signature and consider a planar
matchgate for it.  Consider the $n$ input/output nodes.  If
$\specific_i=1$, add a new edge and a new node to input/output node
$i$.  Move the $i$-th input/output node to the new node.  This has the
effect of switching the value of the $i$-th input bit.  Finally, add
two new nodes with an edge between them, and weight the edge by
$\beta^{-1}$.  The standard signature of the resulting planar
matchgate calculates $g$.  On the other hand, given a planar matchgate
for $g$, we can repeat the process (using $\beta$ instead of
$\beta^{-1}$) and build a planar matchgate for $f$.  Therefore, $f$ is
a standard signature if and only if $g$ is.\eop

In this paper, we are interested in decomposing standard signatures
recursively.  Recall that $\uf_0$ and $\uf_1$ are obtained by fixing
the last bit of a function $f$.  We will repeatedly use the following
fact:

\begin{lemma} \label{lem:f0}
If $f$ is an $(n+1)$-bit standard signature, then $\uf_0$ and $\uf_1$ are
standard signatures.
\end{lemma}

\proof Consider a planar matchgate for $f$.  Let $v$ be the $(n+1)$-st
input/output node.  Consider a new planar matchgate that is identical,
except that $v$ is no longer labelled as an input/output node.  This
planar matchgate calculates $\uf_0$; if we add a new node $v'$ and a
new weight one edge between $v$ and $v'$, the resulting planar
matchgate calculates $\uf_1$.\eop

\subsection{Standard Signatures via Pfaffians} \label{subsec:pfaff}
The determinant of a matrix over a field $\FF$ is a polynomial in the
entries of the matrix.  In the case of a strongly skew-symmetric
$n\times n$ matrix $M$, this polynomial happens to be square, and the
square root is called the Pfaffian.  (We will define the Pfaffian more
formally in a moment.)  If we remove a set of rows and matching
columns from $M$ and calculate the determinant, we produce an object
called a principal minor; there are $2^{n}$ principal minors.  We can
think of this operation (converting a matrix into one of its principal
minors) as a function from $V^n\rightarrow \FF$, where the $i$-th bit
of the input tells us whether or not to delete the $i$-th row and
column.

Suppose, instead of taking the determinant of these submatrices, we
take the Pfaffian.  This will give us another function
$f:V^n\rightarrow \FF$, a sort of square root of the principal minors.
In~\cite{cai3} and~\cite{cai4}, Cai and Choudhary prove that $f$ is a
normalized standard signature; even more amazingly, as we let $M$ vary
over all strongly skew-symmetric matrices over $\FF$, we produce
\emph{all} the normalized standard signatures.

We now state the previous facts and observations more formally.  Let
$m(i,j)$ be the entry of $M$ in the $i$-th row and $j$-th column.  A
matrix $M$ is \emph{strongly skew-symmetric} if $m(i,j)=-m(j,i)$ for
all $i,j$, and $m(i,i)=0$ for all $i$.  (Strong skew-symmetry only
differs from skew-symmetry when the field has characteristic two.)
Note that the set of strongly skew-symmetric matrices can be viewed as
$\FF^{n(n-1)/2}$, since we can determine $M$ by specifying $n(n-1)/2$
entries.

The Pfaffian of an $n\times n$ strongly skew-symmetric matrix $M$ is
defined as zero if $n$ is odd, and one if $n=0$.  If $n=2k$ is a
positive even number, then we define the Pfaffian of $M$ as follows.
Suppose we pair up all the numbers between $1$ and $n$, producing $k$
pairs.  We can encode such a pairing with a permutation that has the
following two properties:
\begin{equation}\label{eq:pf1}
\pi(1)<\pi(2), \pi(3)<\pi(4), ..., \pi(n-1)<\pi(n)
\end{equation}
and
\begin{equation}\label{eq:pf2}
\pi(1)<\pi(3)<\pi(5)<\cdots <\pi(n-1)
\end{equation}
We then view $(\pi(2i-1),\pi(2i))$ as paired numbers for $i=1,...,k$.

Let $\epsilon_\pi$ be the sign of the permutation, i.e.\
$\epsilon_{\pi}=1$ if we can produce $\pi$ from the identity
permutation by composing an even number of transpositions, and
$\epsilon_{\pi}=-1$ otherwise.  Then
\[\pf(M)=\sum_{\pi}\epsilon_{\pi} \prod_{j=1}^k m(\pi(2j-1),\pi(2j)) \]
where the sum runs over permutations $\pi$ satisfying
the inequalities in Formulas~\ref{eq:pf1} and~\ref{eq:pf2}.

There is an alternate definition of $\epsilon_\pi$ that can be useful.
Suppose that we have two pairs of integers $i<j$ and $k<l$, and
suppose that $i<k$.  We say that the two pairs \emph{overlap} if
$i<k<j<l$.  Suppose we consider all the pairs defined by $\pi$.  If
there are an odd number of overlapping pairs, then $\epsilon_\pi=-1$;
otherwise, $\epsilon_\pi=1$.

For $x=(x_1\cdots x_n)\in V^n$, let $M_x$ be the submatrix of $M$
obtained by removing row $i$ and column $i$ from $M$ if $x_i=0$.
Then define $f_M:V^n\rightarrow \FF$ by
\[f_M(x)=\pf(M_x)\]
Cai and Choudhary showed that the set of such functions are precisely
the normalized standard signatures.  Let us state this result formally.

\begin{theorem}[Cai and Choudhary]
Let $\mathcal{M}$ be the set of strongly skew-symmetric $n\times n$ matrices
over a field $\FF$.  Then
\[\{f_M\,|\,M\in\mathcal{M}\} = B_n\]
\end{theorem}

\proof See~\cite{cai3} and~\cite{cai4}.\eop

There is a common method of calculating a determinant by recursively
combining minors. We mention a Pfaffian version of the same thing.

\begin{lemma}
Let $M$ be an $(n+1)\times (n+1)$ strongly skew-symmetric matrix.
Let $\specific=1\cdots 1\in V^{n+1}$.  Then
\begin{eqnarray}
\pf(M) &=& \pf(M_{\specific})\\
\label{eq:pf3}
&=&\sum_{i=1}^n (-1)^{i-1}m(i,n+1)\pf(M_{\specific+e_{i}+e_{n+1}}) \nonumber
\end{eqnarray}

Suppose that there are
$s$ non-zero bits in $x$, and let $p_1,...,p_s$ be the
positions of those bits, in order.  Then
\begin{eqnarray}
\label{eq:pf4}
\pf(M_x) &=& \sum_{i=1}^s (-1)^{i-1}m(p_i,n+1)\pf(M_{x+e_{p_i}+e_{n+1}})\\
\label{eq:pf5}
&=& \sum_{i=1}^n x_i(-1)^{|x|_1^{i-1}}m(p_i,n+1)\pf(M_{x+e_{i}+e_{n+1}})
\end{eqnarray}
\end{lemma}

\proof Equation~\ref{eq:pf3} is standard (see, e.g.~\cite{galbiati});
it can be proved by using the ``overlapping pairs'' definition of
$\epsilon_\pi$.

Equation~\ref{eq:pf4} follows by simply applying Equation~\ref{eq:pf3}
to the submatrix defined by the rows and columns specified by $x$.

Equation~\ref{eq:pf5} follows from Equation~\ref{eq:pf4}, since the
terms in the sum corresponding to irrelevant rows are zeroed out by the
$x_i$ terms, and the $(-1)^{|x|_1^{i-1}}$ term alternates signs at every
non-zero bit in $x$.\eop

For a fixed $n$, we can expand the Pfaffian as a multivariate
polynomial and write down a parameterized expression for the general
form of a normalized standard signature.  The number of terms in the
longest polynomial is of size $O(\sqrt{n!})$, but for small $n$ this
size is manageable.  To see the case of $n=6$ bits, please refer to
Appendix~\ref{app:6bit}.

\subsection{Standard Signatures via Algebraic Varieties} \label{subsec:alg}

The Pfaffian definition of a standard signature above is quite
powerful, and illuminates other interesting structural features of the
standard signatures.  It allows us to describe the set of $n$-bit
standard signatures as an algebraic variety in $\FF^{(2^n)}$.  In
other words, $f$ is a standard signature if and only if the set of
outputs $f(0\cdots 00), f(0\cdots 01),..., f(1\cdots 1)$ satisfy a
collection of polynomial (in fact quadratic) equalities.

We proceed with this alternate definition.  A function $f:V^n
\rightarrow \FF$ is a standard signature if and only if it satisfies
the following two classes of constraints:
\begin{itemize}
\item
First, there is a \emph{Parity Constraint}: $f$ must be an even parity
or odd parity function.
\item Second, there are the \emph{Matchgate
    Identities},\label{def:matchgate.id} also known as the
  \emph{useful Grassmann-Pl\"{u}cker equations}.  Let $p$ be an
  $n$-bit string.  (The ``$p$'' stands for ``position vector''.)  Let
  $L=|p|$.  Let $p_1,...,p_L$ be the positions of the $L$ non-zero
  bits of $p$, in order.
Then for all $\alpha, p\in V^n$, the following equation holds:
\[\sum_{i=1}^L (-1)^i f(\alpha+e_{p_i})f(\alpha+e_{p_i}+p)=0\]
\end{itemize}
The equivalence of these constraints with the Pfaffian definition of a
standard signature was proved by Cai and Choudhary in~\cite{cai3}
and~\cite{cai4}.  We can now prove a few more lemmas.  First, remember
that polynomial images of affine spaces are not necessarily algebraic
varieties (see e.g.\ the exercises in Chapter 3, Section 3
of~\cite{cox}).  In the case of normalized standard signatures, however,
we are lucky:

\begin{lemma} \label{lem:isomorph}
The set $B_n$ of normalized standard signatures is an algebraic
variety isomorphic to $\FF^{n(n-1)/2}$. If $\FF$ is an infinite field,
then $B_n$ has dimension $n(n-1)/2$.
\end{lemma}

\proof Since $A_n$ is an algebraic variety, we can intersect it with
$f(0\cdots 0)=1$ and conclude that $B_n$ is an algebraic variety.

Now, we turn to the isomorphism.  First, since the Pfaffian is a
polynomial in the entries of the matrix $M$, there exists a map
$K:\FF^{n(n-1)/2}\rightarrow \FF^{(2^n)}$ that is surjective on $B_n$.
Next, fix $a<b\leq n$.  Suppose that $\specific=(\specific_1\cdots
\specific_n)$, where $\specific_i=1$ iff $i=a$ or $i=b$.  Then note
that $\pf(M_{\specific})=m(a,b)$.  Therefore, if we project the
coordinates corresponding to weight two codewords, we get a map $K':
B^n\rightarrow \FF^{n(n-1)/2}$ that recovers $M$.  Note that $K'\circ
K$ is the identity in $\FF^{n(n-1)/2}$, and $K\circ K'$ is the
identity on $B^n$.  Therefore, $B_n$ is isomorphic (as an algebraic
variety) to $\FF^{n(n-1)/2}$, and hence they share the same dimension.
If $|\FF|$ is infinite, $\FF^{n(n-1)/2}$ is $n(n-1)/2$
dimensional.\eop

Suppose we take a matchgate $G$ and let the edge weights vary.  Each
choice of edge weights will define a standard signature.  Let $J_G$ be
the collection of such standard signatures, viewed as a subset of
$\FF^{(2^{n})}$.  Then the following lemma holds:

\begin{lemma} \label{lem:J}
Assume that our field $\FF$ is infinite.  Suppose that $G$ is an
$n$-bit planar matchgate.  Suppose that the underlying planar graph of
$G$ has $X$ nodes and $E$ edges.  Then the set $J_G$ is contained in
an algebraic variety of dimension at most $E$.
\end{lemma}


\proof Given a weighted $X$ node planar graph, we can calculate the
sum of all its weighted perfect matchings using the FKT Theorem
(see~\cite{kasteleyn}).  This theorem expresses the sum as the
Pfaffian of a particular $X\times X$ matrix $M$, namely a polynomial
in the edge weights.

If we consider all the $2^{X}$ principal submatrices of the planar
graph, each one corresponds to removing or including a particular node
in the graph (not just the input/output nodes).  The underlying planar
graph forces some of the entries of the matrix to be zero.  If we
ignore that restriction, we have exactly described the set of
normalized standard signatures on $X$ bits.  From
Lemma~\ref{lem:isomorph}, this object is an algebraic variety in
$\FF^{(2^X)}$.  We will now restrict this variety to recover $J_G$.

For each edge $e_{i,j}$ that does not appear in the underlying graph,
we set matrix entries $m(i,j)=m(j,i)=0$.  This results in an
intersection of algebraic varieties, so adding these constraints for
all the missing edges gives us another algebraic variety $P$.  
Since $P$ is parameterized by $E$ variables over $\FF$, it follows that 
$\dim(P)\leq E$.

We are interested in projecting $P$ down to the $2^n$ variables (where
we are only allowed to remove rows and columns corresponding to the
input/output nodes from $M$).  We can now use polynomial
implicitization (see Chapter 3, Section 3, Theorem 1 of~\cite{cox}) to
find the smallest variety $P'$ in $\FF^{(2^n)}$ containing the
projection.  (Note that this theorem assumes that $\FF$ is infinite.)
We construct $P'$ by eliminating variables (i.e.\ intersecting
ideals), so $\dim(P')\leq \dim(P)\leq E$. This establishes our
theorem.\eop

\subsection{Standard Signatures via Recursion} \label{subsec:recursion}

We will present a recursive definition of a standard signature which
makes no explicit reference to Pfaffians or planar matchgates.  We
begin by defining a new set of functions.  Suppose we are given a
non-zero function $f:V^n\rightarrow \FF$ and we choose a base point
$\specific$ such that $f(\specific)\neq 0$.  (We will see in
Corollary~\ref{cor:basepoint} that the choice of base point is
irrelevant for standard signatures; for now, let us choose $\specific$
to be the lexicographically smallest $x$ such that $f(x)\neq 0$.)  Let
us define the \emph{shift basis} functions $s_i^f$ (where $i=1,...,n$)
as
\[s_i^f(x)=
\left\{ 
\begin{array}{ll}
  0 & x_i=\specific_i \\
  (-1)^{|x+\specific|_1^{i-1}}f(x+e_i) & x_i \neq \specific_i\\
\end{array}
\right.
\]
Next let us define the \emph{shift set} as the set of functions formed
by linear combinations of the shift basis functions, i.e.\
\[\SSS_f = \left\{\left.\sum_{i=1}^n \lambda_i s_i^f \, \right| 
\, \lambda_i \in \FF\right\}\]
Note that the elements of the shift set all have the opposite parity as
$f$.

We point out two properties of the shift set.
\begin{lemma} \label{lem:dim}
The shift basis functions for $f$, viewed as vectors over
$\FF^{(2^{n})}$, are linearly independent (i.e.\ they actually form a
basis for $\SSS_f$).  Therefore, $\SSS_f$ can be viewed as an $n$
dimensional subspace of $\FF^{(2^n)}$.
\end{lemma}

\proof 
Notice that 
\[s_i(\specific+e_j)=
\left\{ 
\begin{array}{ll}
  0 & i\neq j \\
  \pm f(\specific)\neq 0 & i=j \\
\end{array}
\right.
\]
so $s_i^f(\specific + e_j)$ is non-zero if and only if $i=j$.  It
follows that the $s_i^f$ are linearly independent, and hence that the
shift set $\SSS_f$ has dimension $n$.\eop

We can now introduce our new definition:
\begin{theorem} \label{thm:main}
The set of normalized standard signatures can be defined recursively:
\begin{equation} \label{eq:norm1}
B_{n+1} = \left\{ f:V^{n+1}\rightarrow \FF \,\,|\,\, \uf_0\in B_n
\mbox{ and } \uf_1\in \SSS_{\uf_0}\right\}
\end{equation}
 The set of all strictly odd or strictly even standard signatures can
 be similarly defined:
\begin{eqnarray}
\label{eq:norm2}
A_{n+1}^{odd} &=&
\left\{ f:V^{n+1}\rightarrow \FF \,\,|\,\, \uf_0\in A_n^{odd}
\mbox{ and } \uf_1\in \SSS_{\uf_0}\right\} \\
&& \cup
\left\{ f:V^{n+1}\rightarrow \FF \,\,|\,\, \uf_0\equiv 0
\mbox{ and } \uf_1\in A_n^{even}\right\}\nonumber 
\end{eqnarray}
\begin{eqnarray}
\label{eq:norm3}
A_{n+1}^{even} &=&
\left\{ f:V^{n+1}\rightarrow \FF \,\,|\,\, \uf_0\in A_n^{even}
\mbox{ and } \uf_1\in \SSS_{\uf_0}\right\} \\
&& \cup
\left\{ f:V^{n+1}\rightarrow \FF \,\,|\,\, \uf_0\equiv 0
\mbox{ and } \uf_1\in A_n^{odd}\right\} \nonumber
\end{eqnarray}

\end{theorem}

\proof We start by proving Equation~\ref{eq:norm1}.  First,
Lemma~\ref{lem:f0} shows that if $f$ is a standard signature, then
$\uf_0$ is a standard signature.

So, assume that $\uf_0$ is a standard signature.  Recall, from our
Pfaffian definition, that for any normalized standard signature
$f:V^{n+1}\rightarrow \FF$, there exists some strongly skew-symmetric
matrix $M$ such that
\[f(x_1\cdots x_{n+1})=\pf(M_{x_1\cdots x_{n+1}})\] Now, suppose that
bit $x_{n+1}=1$, and let $\ux=x_1\cdots x_n$.  

Recall Equation~\ref{eq:pf5}:
\[\pf(M_x)= \sum_{i=1}^n x_i(-1)^{|x|_1^{i-1}}m(p_i,n+1)\pf(M_{x+e_{i}+e_{n+1}})\]
Expressing this in terms of our function $f$, this equation becomes
\begin{eqnarray*}
f(x)&=&\sum_{i=1}^n x_i(-1)^{|x|_1^{i-1}}m(p_i,n+1)f(x+e_{i}+e_{n+1})\\
&=&\sum_{i=1}^n m(p_i,n+1)s_i^{\uf_0}
\end{eqnarray*}
where $s_i^{\uf_0}$ is a shift basis function (with base point
$0\cdots 0$).

Finally, since the $(n+1)$st bit of $x=1$, we can write
\[f(x)=\uf_1(\ux)=\sum_{i=1}^n m(i,n+1)s_i^{\uf_0}(\ux)\]

In other words, the set of valid $\uf_1$ is exactly $\SSS_{\uf_0}$.  
In other words,
\[B_{n+1} = \left\{ f:V^{n+1}\rightarrow \FF \,\,|\,\, \uf_0\in B_n
\mbox{ and } \uf_1\in \SSS_{\uf_0}\right\}\]
which establishes Equation~\ref{eq:norm1}.

Next, let us turn to proving Equation~\ref{eq:norm2}.  Consider $f\in
A_{n+1}^{odd}$ where $\uf_0\neq 0$.  Let $\specific=
\specific_1\cdots\specific_{n+1}\in V^{n+1}$ such that
$f(\specific)\neq 0$ and $\specific_{n+1}=0$.  Let
$g(x)=(1/f(\specific))f(x+\specific)$, i.e.\ $f$ normalized around
$\specific$.  From Equation~\ref{eq:norm1}, we know that $g$ is a
standard signature if and only if $\ug_1\in\SSS_{\ug_0}$.  If we
translate the elements of $\SSS_{\ug_0}$ by adding $\specific$ to the
inputs, notice that the resulting set is exactly $\SSS_{\uf_0}$.
Using Lemma~\ref{lem:normalize}, we can conclude that $f$ is a
standard signature if and only if $\uf_1\in\SSS_{\uf_0}$.  This
establishes the first half of Equation~\ref{eq:norm2}.

On the other hand, if $f\in A_{n+1}$ but $f\equiv 0$, then $f_1 \in
A_{n}$.  We know that $f_1\subseteq A_n$ from
Lemma~\ref{lem:f0}. Conversely, given any $f_1\in A_n$, we can
construct a planar matchgate for $f$ by adding a new disconnected node
and labelling it as input/output node $n+1$.  This establishes the
second half of Equation~\ref{eq:norm2}.

Equation~\ref{eq:norm3} follows symmetrically to
Equation~\ref{eq:norm2}.\eop

Theorem~\ref{thm:main} gives us a recursive procedure to determine if
an $n$-bit function is a standard signature.  If the function is the
constant zero function, it is a standard signature.  Otherwise, we can
normalize it to a function $f$.  We can now check if
$\uf_1\in\SSS_{f_0}$ and if $\uf_0\in B_n$.  The first condition can
be checked by linear algebra, and the second condition can be checked
recursively.

Finally, we justify our earlier comment about the irrelevance of our
choice of base point for normalization.

\begin{corollary} \label{cor:basepoint}
If $g$ is a non-zero standard signature, then the set $\SSS_g$ is
independent of the choice of base point.
\end{corollary}

\proof Suppose we have a standard signature $f$ where there exist two
base points around which we can normalize $\uf_0$ (i.e.\ there exist
$b\neq c$ such that $b_{n+1}=c_{n+1}=0$, where $f(b)\neq 0$ and
$f(c)\neq 0$.)  These different definitions of ``normalization''
produce two possibly different sets $\SSS_{\uf_0}$ and
$\SSS'_{\uf_0}$.  If we applied the proof of Theorem~\ref{thm:main} to
each case, we would conclude that $f$ is standard signature iff
$\uf_1\in \SSS_{f_0}$ iff $\uf_1\in\SSS_{\uf_0}'$.  Therefore,
$\SSS_{\uf_0}=\SSS_{\uf_0}'$.  Now, for any $g\in A_n$ there exists an
$(n+1)$-bit standard signature $f$ such that $\uf_0=g$.  Therefore,
$\SSS_g$ is independent of the choice of base point.  \eop

\section{Consequences of Recursion} \label{sec:hints}
\subsection{Counting Standard Signatures} \label{subsec:count}
Over a finite field $\FF$, there are only finitely many $n$-bit
standard signatures for any fixed $n$.  In other words, $|A_n|$ is
finite.  The recursive structure described in Theorem~\ref{thm:main}
allows us to find a formula to count $|A_n|$.

\begin{corollary} \label{cor:count}
If we are operating in a finite field $\FF$, where $|\FF|=s$, then we
can calculate the cardinality of the set of normalized standard
signatures, odd parity standard signatures, and general standard
signatures:
\begin{eqnarray} \label{eq:count_norm}
\left| B_n \right| &=& \prod_{i=1}^{n-1}  s^{i}
= s^{n(n-1)/2}\\
\label{eq:count_odd}
\left| A_n^{odd} \right| &=& \left(s-1\right)\prod_{i=1}^{n-1}
\left(s^{i}+1\right)\\
\label{eq:count}
\left| A_n \right| &=& 
1+2\times\left[\left(s-1\right)\prod_{i=1}^{n-1}  \left(s^{i}+1\right)\right] 
\end{eqnarray}
where we interpret the empty product $\prod_{i=1}^0$ as evaluating to
one.
\end{corollary}

\proof We first consider Equation~\ref{eq:count_norm}.  For any
$\uf_0\in B_n$, $\uf_1$ can be chosen freely from $\SSS_{\uf_0}$
Lemma~\ref{lem:dim} shows that $\SSS_{\uf_0}$ is $n$-dimensional, so
\[\left|\SSS_{\uf_0}\right|=s^n\]
regardless of which particular (non-zero) $\uf_0$ we pick.  If $n>1$,
then Theorem~\ref{thm:main} implies that
\begin{eqnarray*}
|B_{n+1}|&=&
\left|\left\{ f:V^{n+1}\rightarrow \FF \,\,|\,\, \uf_0\in B_n
\mbox{ and } \uf_1\in \SSS_{\uf_0}\right\}\right|\\
&=&|B_n|\times s^n
\end{eqnarray*}
Since $|B_1|=1$, Equation~\ref{eq:count_norm} follows by induction.

Next, consider Equation~\ref{eq:count_odd}.  
If $n>1$, then note that Equation~\ref{eq:norm2} of Theorem~\ref{thm:main}
is a disjoint union of two sets.  Therefore,
\begin{eqnarray*}
|A_{n+1}^{odd}| &=&\left|
\left\{ f:V^{n+1}\rightarrow \FF \,\,|\,\, \uf_0\in A_n^{odd}
\mbox{ and } \uf_1\in \SSS_{\uf_0}\right\} \right.\\
&& \left. \cup
\left\{ f:V^{n+1}\rightarrow \FF \,\,|\,\, \uf_0\equiv 0
\mbox{ and } \uf_1\in A_n^{even}\right\}
\right|\\
&=&
\left| \left\{ f:V^{n+1}\rightarrow \FF \,\,|\,\, \uf_0\in A_n^{odd}
\mbox{ and } \uf_1\in \SSS_{\uf_0}\right\} \right| \\
&& +\left|
\left\{ f:V^{n+1}\rightarrow \FF \,\,|\,\, \uf_0\equiv 0
\mbox{ and } \uf_1\in A_n^{even}\right\}
\right|\\
&=& \left|A_n^{odd}\right|\times s^n + \left|A_n^{odd}\right|\\
&=& (s^n+1)\left|A_n^{odd}\right|
\end{eqnarray*}
By Lemma~\ref{lem:a1}, $\left|A_1^{odd}\right|=s-1$.  Therefore,
by induction, we have proved Equation~\ref{eq:count_odd}.

Finally, we turn to Equation~\ref{eq:count}.  By
Lemma~\ref{lemma:involute}, 
$\left|A_n^{odd}\right|=\left|A_n^{even}\right|$.  If we account
for the zero function, we can conclude that
\begin{eqnarray*}
\left|A_n\right|&=&1+2\times\left|A_n^{odd}\right|\\
&=&1+2\times\left[\left(s-1\right)\prod_{i=1}^{n-1}  
\left(s^{i}+1\right)\right]
\end{eqnarray*}

\subsection{Asymptotics of $\left|A_n\right|$} \label{subsec:asymptotics}
If we want to evaluate $|A_n|$ or $|A_n^{odd}|$ for small $s$ and $n$,
we can just plug in to Equations~\ref{eq:count_odd} or~\ref{eq:count}.
However, we might also be interested in the behavior for fixed $s$ as
$n$ grows.

In order to study this regime, we will introduce the (partial)
function $\gamma:\CC\rightarrow \CC$, where
\[\gamma(x)=\prod_{i=1}^{\infty}(1+(1/x)^i)\]
It is not a priori clear that $\gamma$ converges.  However, if we
expand in $1/x$, then $\gamma$ is the generating function for the
number of ways of partitioning a set into unequal parts.  We can then
use the following lemma:
\begin{lemma}
If $x\in\CC$ lies outside the unit circle, then $\gamma(x)$ converges.
\end{lemma}

\proof A proof can be found in~\cite{apostol}, Section~14.4.\eop

We are interested in integer values of $x$ where $x\geq 2$, so
$\gamma(x)$ will converge.  We can now express the asymptotics of
$|A_n|$ more precisely.

\begin{theorem}
Suppose we are operating on a finite field $\FF$ of size $|\FF|=s$.
Then
\begin{eqnarray*}
\lim_{n\rightarrow \infty} \frac{|A_n^{odd}|}{s^{n(n-1)/2+1}} &=&
\gamma(s) \\
\lim_{n\rightarrow \infty} \frac{|A_n|}{s^{n(n-1)/2+1}} &=&
2\gamma(s)
\end{eqnarray*}
Therefore, the growth rate is
\begin{eqnarray*}
|A_n^{odd}|&=&\Theta\left(s^{n(n-1)/2 +1}\right)\\
|A_n|&=&\Theta\left(s^{n(n-1)/2 +1}\right)
\end{eqnarray*}
\end{theorem}

\proof
From Theorem~\ref{cor:count}, we can write
\begin{eqnarray*}
\left| A_n^{odd} \right| &=& \left(s-1\right)\prod_{i=1}^{n-1}
\left(s^{i}+1\right)\\
 &=& \left(s-1\right)s^{n(n-1)/2}\prod_{i=1}^{n-1}
\left(s^{i}+1\right)/s^i\\
 &=& \left(s-1\right)s^{n(n-1)/2}\prod_{i=1}^{n-1}
\left(1+1/(s^i)\right)
\end{eqnarray*}
Therefore,
\begin{eqnarray*}
\lim_{n\rightarrow \infty} \frac{|A_n^{odd}|}{s^{n(n-1)/2+1}} &=&
\lim_{n\rightarrow \infty} \frac{|A_n^{odd}|}{(s-1)s^{n(n-1)}} \\
&=&
\lim_{n\rightarrow \infty} \prod_{i=1}^{\infty}\left(1+(1/s)^i\right)\\
&=&
\gamma(s)
\end{eqnarray*}
It follows that $|A_n^{odd}|=\Theta\left(s^{n(n-1)/2 +1}\right)$.

Since $|A_n|=1+2|A_n^{odd}|$, the results on $|A_n|$ follow.\eop

In practice, the product form for $\gamma$ converges somewhat slowly.
However, there is a trick for evaluating $\gamma$ more efficiently.
Recall Euler's Pentagonal
Formula (see~\cite{apostol}): for any $|\sigma|<1$,
\[\prod_{i=1}^\infty\left(1-\sigma^r\right) = 
\sum_{i=-\infty}^{\infty}(-1)^i\sigma^{\omega(i)}\] where
$\omega(i)=(3i^2-i)/2$.  The sum formulation converges much more
rapidly.  If we let $\sigma=1/s$, we can write
\begin{eqnarray}
\nonumber
\gamma(1/\sigma)&=&
\prod_{i=1}^{\infty}(1+\sigma^{i})\\
\label{eq:euler1}
&=&\prod_{i=1}^{\infty}\frac
{1-\sigma^{2i}}{1-\sigma^{i}}\\
\label{eq:euler2}
&=&\frac
{\prod_{i=1}^{\infty}(1-\sigma^{2i})}{\prod_{i=1}^{\infty}(1-\sigma^{i})}\\
\nonumber
&=&\frac{\sum_{i=-\infty}^{\infty}(-1)^i\sigma^{2\omega(i)}}
{\sum_{i=-\infty}^{\infty}(-1)^i\sigma^{\omega(i)}}
\end{eqnarray}
Since the products are infinite, the step from
Equation~\ref{eq:euler1} to Equation~\ref{eq:euler2} requires
justification, but it is straightforward.

It now becomes computationally simple to calculate $\gamma(s)$ to high
precision; here is a table for a few values:

\begin{center}
\begin{tabular}{|l|l|}
\hline
$s=|\FF|$ & $\gamma(s)$\\
\hline
2 & 2.384231 \\ 
3 & 1.564934 \\
4 & 1.355910 \\
5 & 1.260501 \\
7 & 1.170149 \\
8 & 1.145129 \\
9 & 1.126565 \\
\hline
\end{tabular}
\end{center}

So, for instance, for large $n$, there are about
\[2\gamma(2)2^{n(n-1)/2+1}=4.768\times 2^{n(n-1)/2 +1}\] 
$n$-bit standard signatures over $\FF_2$.  These calculations will
also enable us to calculate the table of probabilities in
Subsection~\ref{subsec:sparse}.

\subsection{Bounds on Planar Matchgate Sizes} \label{subsec:matchgate.size}

If we are given an $n$-bit standard signature, by definition there
exists some planar matchgate that computes it.  However, it is not a
priori clear how large the planar matchgate must be to simulate the
standard signature.  An upper bound of size $O(n^4)$ on the number of
nodes and edges has been constructed by Li and Xia (see Theorem 3.3
in~\cite{li}), and in Appendix~\ref{app:matchgate}, we mention a
recursive construction that would require $O(n^3)$ nodes and edges.
However, these bounds are both beaten by Cai and Choudhary's original
constructions in~\cite{cai3} and~\cite{cai4}, which establish an
$O(n^2)$ upper bound on the number of nodes and edges
required.\footnote{In fact, if we apply the switch planar matchgates
  in Appendix~\ref{app:matchgate} to Cai and Choudhary's construction,
  we can produce a planar matchgate for an $n$-bit standard signature
  on any field that uses at most $20n(n-1)+n+2$ nodes.  For fields of
  characteristic two, $7n(n-1)+n+2$ nodes suffice.}

In this subsection, we present a matching lower bound showing that the
$O(n^2)$ upper bound is tight.

\begin{theorem} \label{thm:node.lower}
There exist standard signatures that can only be represented on graphs
with at least $\Omega(n^2)$ nodes.  More specifically, there exist
standard signatures that require $X$ nodes, where
\[X + O(\log(X)) > n^2/16.015 - O(n\log (n))\]
\end{theorem}

\proof First, suppose that $\FF$ is an infinite field.  Suppose we
choose:
\begin{itemize}
\item
 an unweighted planar graph with at most $X$ nodes, where $X\geq n$,
 along with
\item
some planar embedding for the graph, and 
\item
a choice of $n$ input/output nodes on the outer face.
\end{itemize}
We will call such an object a \emph{stripped matchgate}, since we have
stripped off the edge weights.  If we take a stripped matchgate and
add edge weights, we get a planar matchgate.

We will consider two planar embeddings to be isomorphic if they
produce the same set of nodes on the outer face, in the same order.
Note that there are only finitely many non-isomorphic planar
embeddings for any graph.  Since the other properties of a stripped
matchgate are also finitary, it follows that that there are only
finitely many stripped matchgates with non-isomorphic planar
embeddings.  Let $\mathcal{G}$ be a set of planar matchgates
representing each of the possible stripped matchgates with
non-isomorphic planar embeddings; our comments above show that
$|\mathcal{G}|$ is finite.

If $G$ is a representative planar matchgate then recall from
Subsection~\ref{subsec:alg} that $J_G$ is the set of all standard
signatures sharing the same stripped matchgate.

Suppose our graph has $E$ edges.  Since our graph is planar, $E\leq
3X$.  Lemma~\ref{lem:J} shows that $J_G$ is contained in an algebraic
variety $P_G$ with $\dim(P_G)\leq E\leq 3X$ for any $G$.  Therefore,
the set of standard signatures definable on graphs with at most $X$
nodes is contained in a finite union of varieties:
$\cup_{G\in\mathcal{G}} P_G$.  This finite union is itself a variety;
since each component has dimension at most $E$, the union has
dimension at most $E\leq 3X$.

However, recall from Lemma~\ref{lem:isomorph} that $B_n$ is
also an algebraic variety, and $\dim(B_n)=n(n-1)/2$.  Therefore,
if $3X < n(n-1)/2$, then
\[\dim(B_n) =n(n-1)/2 > 3X \geq \dim(\cup_{G\in\mathcal{G}} P_G)\]
Therefore,
\[B_n \not\subseteq \cup_{G\in\mathcal{G}} P_G\]
and hence
\[B_n \not\subseteq \cup_{G\in\mathcal{G}} J_G\]
Therefore, there exist standard signatures in $B_n$ (and thus $A_n$)
that require at least $n(n-1)/6=n^2-O(n)=\Omega(n^2)$ signatures to
represent them.

Next, suppose that $\FF$ is a finite field.  Roughly speaking, we will
repeat the argument above, but the finiteness of the number of planar
matchgates is no longer sufficient-- we need to count the number of
planar matchgates explicitly, which is a more delicate operation.

Suppose we consider a planar matchgate with underlying (weighted)
graph $G$ on $X$ nodes.  We are going to represent the planar
matchgate as a planar graph on $X+1$ nodes with certain special
labels.  We proceed as follows: we take $G$ and add a new node $v$.
We label this node as ``extra''.
We add an edge from $v$ to each of the input/output nodes, and give
the new edges weight one.  We label each of the $n$ input/output nodes
by a distinct number from 1 to $n$, namely the number of the node.

Let $T_X$ be the set of labelled planar graphs with a node labelled
``extra'', which has $n$ neighbors, each labelled with a distinct
number between 1 and $n$.  (So the elements of $T_X$ are graphs with
$X+1$ nodes.)  Note that $T_X$ is larger than the set of planar
matchgates, because we are not enforcing the input/output nodes to be
on the outer face of the graph.  However, every different $X$ node
matchgate maps to a distinct one of these labelled planar graphs, so
by counting $|T_X|$, we will get an upper bound on the number of
standard signatures that can be represented with $X$ node graphs.
Note also that we are counting planar graphs, not planar embeddings (a
different embedding of the same matchgate will produce the same
standard signature, assuming that the input/output nodes are still on
the outer face, and we orient the embedding to make the node labels
run anti-clockwise.)

The reader may wonder how we can add the ``extra'' node $v$ and its
edges and be confident that our graph remains planar.  The
input/output nodes all lie on the outer face of some planar
embedding; therefore, it is possible to place a node in the outer face
and attach it to all the input/output nodes without crossing any
edges.

Suppose we are given a planar matchgate with $X-2Y$ nodes.  Then we
can add disconnected 2-node subgraphs with edges of weight 1 at will
without changing the standard signature.  If we add $Y$ of those
subgraphs, we build a planar matchgate with $X$ nodes.  Therefore, all
standard signatures representable on planar matchgates with $X-2Y$
nodes are representable on planar matchgates with exactly $X$ nodes.

Therefore, all standard signatures on planar matchgates with at most
$X$ nodes can be represented by unique elements of $T_{X}$ or $T_{X-1}$.

We now need to determine the size of $T_X$.  Planarity is a very
restrictive condition on a graph; there at most $2^{5.007X + O(\log
  X)}$ planar graphs with $X$ (unlabelled) nodes
(see~\cite{bonichon}).  There are at most $3X$ edges on a planar
graph, so we have at most $(s-1)^{3X}$ labellings.  There are
$X$ possible choices for the ``extra'' node.  The neighbors of the
extra node are all labelled by distinct numbers between $1$ and $n$,
so there are $n!$ possible numberings.
Therefore,
\[|T_X|\leq (X+1)(n!)(s-1)^{3(X+1)}2^{5.007(X+1) + O(\log(X+1))}\]
Therefore,
\begin{eqnarray*}
|T_{X}|+|T_{X-1}|&\leq& 2(X+1)(n!)(s-1)^{3(X+1)}2^{5.007(X+1) + O(\log(X+1))}
\end{eqnarray*}
Bringing all the terms into the exponent and absorbing extraneous ones
into the $O(\log(X))$ term, (and remembering that $n!=2^{n\log_2
  (n/e)+ O(\log(n))}$), we can rewrite this as
\[|T_X|+|T_{X-1}|\leq 2^{5.007X +3X\log_2(s-1)+ n\log_2(n/e) + O(\log(X))}\]

However, we know that there are
\[1+2(s-1)\prod_{i=1}^{n-1}(s^i+1)>s^{n(n-1)/2}=2^{n(n-1)(\log_2(s))/2}\]
$n$-bit standard signatures.  Therefore, in order to express all
these standard signatures, we need $X$ to be at least large enough that
\[2^{5.007X +3X\log_2(s-1)+ n\log_2(n/e) + O(\log(X))}
> 2^{\log_2(s)n(n-1)/2}\]
Comparing exponents, we therefore need 
\[5.007X +3X\log_2(s-1)+ n\log_2(n/e) + O(\log(X))> \log_2(s)n(n-1)/2\]
Replacing $(s-1)$ by $s$ on the left hand side and solving for $X$,
we get
\[X + O(\log(X)) > \frac{n(n-1)-\frac{2n\log_2(n/e)}{\log_2(s)}}
    {6+(10.014/\log_2(s))} + O(\log(n))\]
So, there must exist some standard signature that requires at least
\[\frac{n(n-1)-\frac{n\log_2(X)}{2\log_2(s)}}
    {6+(10.014/\log_2(s))} + O(\log(n))\]
nodes.  This lower bound is $\Omega(n^2)$, so we have established
the rough bound for the theorem.  To obtain the specific bound,
note that the denominator is maximized when $s=2$, at which point
the denominator becomes $16.014...$.  Conservatively rounding it up
to $16.015$ gives the result.\eop

\subsection{Efficiently Detecting Standard Signatures} 
\label{subsec:complexity}
Suppose we are given a function $f:V^n\rightarrow \FF$, and we would
like to determine if $f$ is a standard signature.  What is the
complexity of deciding that question?

First, let us find a lower bound.  The function $f$ has $2^n$ inputs.
Suppose that $f(0\cdots 0)\neq 0$.  At the very least, we need to
check that all the $2^{n-1}$ odd-parity strings map to zero.
Therefore, deciding if $f$ is a standard signature takes at least
\[2^{n-1}=\Omega(2^n)\]
steps to evaluate.

But how should we actually verify that $f$ is a standard signature?
One reasonable approach would be to use the algebraic variety defining
$A_n$.  Recall that $f$ is a standard signature iff it satisfies the
Parity Constraint and the Matchgate Identities for every $p, \alpha\in
V^n$.  We can verify the Parity Constraint by running through the
output values once and checking for non-zero values, which takes $2^n$
steps.  For the Matchgate Identities, each equation has $n$ terms, and
there are $2^n$ choices for both $p$ and $\alpha$.  Assuming the
Parity Constraint holds, we only need to check the Matchgate
Identities for even parity $p$ and $\alpha$ of opposite parity to $f$.
This approach would take
\[2^n+n2^{2(n-1)}=O(n2^{2n})\]
steps to evaluate.\footnote{If our function $f$ happens to be sparse,
  with only $k$ non-zero values, then we only need to check at most
  $n{{k}\choose{2}}$ Matchgate Identities.  Therefore, we can
  determine if $f$ is a standard signature in only
  $n^2{{k}\choose{2}}$ steps.}

The recursive structure of the standard signatures allows us to use a
much more efficient approach.  The general outline of our technique is
to assume that $f$ is a standard signature.  This assumption lets us
recover a unique fingerprint for $f$ by examining only a small subset
of the output values.  We then use this fingerprint to reconstruct an
actual standard signature $f'$; this reconstruction takes $n2^n$
steps.  Finally, $f$ is a standard signature iff $f\equiv f'$, which
we can check in another $2^n$ steps.  This approach takes only
\[O(n2^n)\]
steps to evaluate.  We now analyze this process more carefully.

\begin{theorem}
Suppose we are given a function $f:V^n\rightarrow \FF$ (that is, we
are given a list of $f(x)$ for all $x\in V^n$, sorted by $x$).  Then
we can determine if $f$ is a standard signature in time $O(n2^n)$.
\end{theorem}

\proof We begin by determining if $f$ is identically zero.  This takes
$O(2^n)$ steps; if $f\equiv 0$ then it is a standard signature, and we
are done.  Otherwise, we will discover a string $\specific\in V^n$
such that $f(\specific)\neq 0$.  Let us normalize our function at
$\specific$ by constructing the new function
$g(x)=(1/f(\specific))f(x+\specific)$.  From
Lemma~\ref{lem:normalize}, $f$ is a standard signature if and only if
$g$ is, so we will henceforth focus on $g$.  Constructing $g$ takes
another $O(2^n)$ steps.

Suppose that we have a standard signature $h$.
Recall from the Pfaffian definition of the standard signature that
there is some matrix $M$ such that
\[h(x)=\pf(M_x)\] Let us use $m(i,j)$ to represent the entry of $M$ in
the $i$-th row and $j$-th column.  Suppose that $x$ has Hamming weight
two, i.e.\ $x=e_i+e_j$, where $i<j$.  Then $M_x$ is a $2\times 2$
matrix of the form
\[\left(\begin{array}{ll}
  0 & m(i,j) \\
  -m(i,j) & 0 
\end{array}\right)\]

In particular,
\[h(x)=\pf(M_x)=m(i,j)=-m(j,i)\]
In other words, the $n(n-1)/2$ weight 2 codewords completely specify
$M$.

So, given $g$, let $M$ be the matrix determined by the value of $g$ on
all the weight-two codewords.  Now that we have $M$, let us construct
a standard signature $h$ from it.  We do this recursively.
Define $h^1(0)=1, h^1(1)=0$.  Define
\[h^j(x_1\cdots x_j) = \left\{ \begin{array}{ll}
  h^{j-1}(x_1\cdots x_{j-1}) & \mbox{if }x_j=0 \\
  \sum_{i=1}^{j-1} -m(i,j)s_i^{h^{j-1}} & \mbox{if }x_j=1
  \end{array}
\right.\] where $s_i^{h^{j-1}}$ is a shift-basis function of
$h^{j-1}$.  Recovering $h^n$ takes 
\[\sum_{i=2}^n(i-1)2^{i-1}=(n-2)2^n + 2=O(n2^n)\]
steps.  From our recursive definition of the standard signatures
(cf.\ the proof of Theorem~\ref{thm:main}), it follows that $h^n$ is a
standard signature, and by construction $h^n(x)=g(x)$ for all
weight-two codewords $x$.

Since each standard signature defines a unique $M$, $g$ is a standard
signature if and only if $g\equiv h^n$.  We can compare their
outputs in $2^n$ steps; they are identical if and only if $g$ (and hence $f$)
is a standard signature.\eop

\subsection{Expected Sparsity} \label{subsec:sparse}

How large is the support of a typical standard signature?  That is, if
we choose $f\in A_n$ ``randomly'', what fraction of the entries are
non-zero?  To put it another way, if we view $f$ as a vector in
$\FF^{(2^n)}$, how sparse is the vector?

For infinite fields, it is not clear which measure we should use to
select our function $f$.  But if $\FF$ is a finite field, it seems
natural to choose $f$ uniformly at random from, say, $A_n^{even}$, and
the problem is well-defined.  It turns out that we can prove a
slightly stronger result-- we can calculate the expected sparsity for
each individual input bit.

\begin{theorem} \label{cor:prob}
Assume we are operating over a finite field $\FF$ of size $s=|\FF|$.
Suppose we choose $f\in A_n^{even}$ uniformly at random, and select
any fixed even parity $n$-bit string $\specific$.  Then
\begin{equation} \label{eq:prob}
\Pr(f(\specific)\neq 0) = \left[\prod_{i=1}^{n-1}(1+s^{-i})\right]^{-1}
\end{equation}
The analogous result also holds for strictly odd parity standard signatures.
\end{theorem}

\proof Let $C_n^{even}\subseteq A_n^{even}$ such that for any $g\in
C_n^{even}$, $g(0\cdots 0)\neq 0$.  Note that if we take an element of
$f$ and divide its outputs by $f(0\cdots 0)$, we obtain a normalized
standard signature in $B_n$.  Each element of $B_n$ is the image of
exactly $s-1$ elements of $C_n$.  Therefore,
\[|C_n|=(s-1)|B_n|=(s-1)\prod_{i=2}^ns^{i-1}\]
Now, if we choose $f\in A_n^{even}$ uniformly at random, notice that
\[\Pr[f(0\cdots 0)\neq 0]=\Pr[f\in C_n^{even}]\]
Since we are selecting functions uniformly, it follows that
\begin{eqnarray*}
\Pr[f(0\cdots 0)\neq 0]&=&\frac{|C_n^{even}|}{|A_n^{even}|}\\
&=&\frac{(s-1)\prod_{i=1}^{n-1}s^{i}}{(s-1)\prod_{i=1}^{n-1}(1+s^i)}\\
&=&\prod_{i=1}^{n-1}\frac{s^i}{1+s^i}\\
&=&\prod_{i=1}^{n-1}\frac{1}{1+s^{-i}}\\
&=&\left[\prod_{i=1}^{n-1}(1+s^{-i})\right]^{-1}
\end{eqnarray*}

Suppose we take the classes of functions above and add $\specific$ to
their inputs (i.e.\ we translate them by $\specific$).  The sizes of
the sets, and thus the probabilities, do not change.  Therefore, we
can conclude that for any fixed $\specific$,
\[\Pr[f(\specific)\neq 0]=
\left[\prod_{i=1}^{n-1}(1+s^{-i})\right]^{-1}\]
as desired.\eop

As a simple consequence, we can calculate the expected sparsity:
\begin{corollary}
If we choose non-zero $f\in A_n$ uniformly at random, then
\begin{eqnarray*}
\mbox{Expected Sparsity}&:=&E\left[\frac{|\{x\,|\, f(x)\neq0\}|}{2^n}\right]\\
&=&\left[\prod_{i=1}^{n-1}(1+s^{-i})\right]^{-1}
\end{eqnarray*}
\end{corollary}

\proof 
\begin{eqnarray*}
\mbox{Expected Sparsity}&:=&E\left[\frac{|\{x\,|\, f(x)\neq0\}|}{2^n}\right]\\
&=&\frac{1}{2^n}E\left[\sum_{x\in V^n}\Pr(f(x)\neq 0)\right]\\
&=&\frac{2^n}{2^n}\left[\prod_{i=1}^{n-1}(1+s^{-i})\right]^{-1}\\
&=&\left[\prod_{i=1}^{n-1}(1+s^{-i})\right]^{-1}
\end{eqnarray*}
as desired.\eop

For a fixed field of size $s$, the probability converges as
$n\rightarrow \infty$:
\begin{eqnarray*}
&&\lim_{n\rightarrow \infty}\left(\mbox{Expected Sparsity of }
A_n^{even}\right)\\
&=&\lim_{n\rightarrow \infty}(\Pr(f(0\cdots 0)\neq 0))\\
&=&\lim_{n\rightarrow \infty}\left[\prod_{i=1}^{n-1}(1+s^{-i})\right]^{-1}\\
&=& 1/\gamma(s)
\end{eqnarray*}
(See Subsection~\ref{subsec:asymptotics} for details and computational
issues.)  We include a table of these limiting probabilities for a few
small fields.  For comparison, we also list the expected sparsity of
an arbitary function $g:V^n\rightarrow \FF$ selected uniformly at
random, which equals $1-(1/s)$.

\begin{center}\label{table:prob}
\begin{tabular}{|l|l|l|} 
\hline
$s=|\FF|$ & $1/\gamma(s)$ & $1-(1/s)$ \\
\hline
2 & 0.419422 & 0.5\\
3 & 0.639005 & 0.666666\\
4 & 0.737512 & 0.75\\
5 & 0.793335 & 0.8\\
7 & 0.854592 & 0.857142\\
8 & 0.873264 & 0.875\\
9 & 0.887654 & 0.888888\\
\hline
\end{tabular}
\end{center}

\subsection{Expressiveness of Holographic Algorithms} 
      \label{subsec:expressiveness}
If our base field $\FF$ is finite, then there are only a finite number
of $n$-bit standard signatures.  In addition to being of intrinsic
interest, the number of standard signatures gives us some intuition
about the expressiveness of a holographic algorithm: the more
signatures, the more expressive the algorithms could possibly be.  We
have found it instructive to compare the relative sizes of a few
classes of functions.

Let $|\FF|=s$.
\begin{itemize}
\item  The number of functions from $V^n$ to $\FF$:
\[s^{\left(2^{n}\right)}\]
\item  The number of functions from $V^n$ to $\FF$ with
even or odd parity:
\[2s^{\left(2^{n-1}\right)}-1\]
\item  The number of standard signatures:
\[|A_n|=1+2\times\left[(s-1)\prod_{i=1}^{n-1}  (s^{i}+1)\right]  
= \Theta\left(s^{n(n-1)/2 + 1}\right)\]
\item The number of symmetric realizable functions (assuming the
  characteristic of the field is odd, and the characteristic doesn't
  divide $n$) on \emph{any} basis of size 1 (not just the standard
  basis):
\[s(s-1)^3(s+3)+1=\Theta(s^5)\]
See Theorem~4.2 in~\cite{cai1} for more details.
\end{itemize}

So, based only on cardinality, we could argue that general functions
are exponentially more expressive than standard signatures, which, in
turn, are exponentially more expressive than symmetric realizable
functions.

\subsection{Cardinality of Self-dual Codes} \label{subsec:self.dual}
It would be extremely interesting to find an isomorphism between the
$n$-bit standard signatures and other, better studied mathematical
objects.  Having an exact count of the number of standard signatures
over various finite fields can facilitate this hunt; if an isomorphic
object exists, it will necessarily have the same cardinality.  Do any
such objects exist?

We can find an example in the world of self-dual codes. Recall that
over $\FF_2$, $|A_n^{odd}|=\prod_{i=1}^{n-1}(2^i+1)$. Surprisingly,
this equals the number of dimension $n$ self-dual codes over $\FF_2$
(i.e.\ self-dual codes in $\FF_2^{2n}$).  Moreover, over $\FF_3$, it
turns out that $|A_n^{odd}|=2\prod_{i=1}^{n-1}(3^i+1)$ equals the
number of dimension $n$ self-dual codes over $\FF_3$!  See Chapter 3
of~\cite{pless} for these results; consider ``type $q^{E}$'' self-dual
codes.

When discussing a self-dual code, we implicitly assume some particular
inner product; for the two results above, we used the Euclidean inner
product.  Frustratingly, if we continue to use the same inner product,
the cardinalities diverge for all other finite fields.  The agreement
over $\FF_2$ and $\FF_3$ seems like a fairly spectacular coincidence,
though.

How can we circumvent this divergence?  We might look for a better
inner product, but no obvious candidates suggest themselves.
(See~\cite{nebe} for a thorough examination of many alternate
possibilities.)  If we stick to the Euclidean inner product, though,
we can match the cardinalities with a little normalization gimmick.
For any $f\in A_n^{odd}$, since $f\not\equiv 0$, there exists a
lexicographically smallest $\specific\in V^n$ such that
$f(\specific)\neq 0$.  Call a standard signature
\emph{semi-normalized} if $f(\specific)=1$.  Let $H_n^{odd}$ be the
set of semi-normalized standard signatures.  Let $\#SD(\FF,n)$ be the
number of $n$ dimensional self-dual codes over $\FF$ with the
Euclidean inner product.  If $|\FF|$ is even, it turns out that
\[|H_n^{odd}|=\#SD(\FF,n) \]
while if $|\FF|$ is odd, then
\[|H_n^{odd}| + |H_n^{even}|=\#SD(\FF,n) \]
Although numerically surprising, the above observations do not suggest
how we might actually construct an isomorphism between the standard
signatures and the self-dual codes.  Until we can build a non-trivial
isomorphism, these cardinality results remain only curiosities.

\appendix
\section{The Six-bit Normalized Standard Signature} \label{app:6bit}
Recall that a normalized standard signature is a standard signature
$f$ where $f(0\cdots0)=1$.  
Since $f$ is an even function, we only need to specify the output for
even-weight inputs; all the odd-weight inputs evaluate to zero.  

As we discussed in Subsections~\ref{subsec:pfaff}
and~\ref{subsec:sparse}, all the outputs can be expressed as
polynomials in the $f(\specific)$, for $\specific$ with Hamming weight
two.  More generally, if $x$ has Hamming weight $2k$, then $f(x)$ can
be expressed in terms of $f(x')$ where $x'$ has weight $2k-2$.  Each
monomial term in a polynomial has coefficient $\epsilon_\pi=\pm 1$.
It is straightforward to show by induction on the weight of the input
string that if $|x|=2k$, then there are
\[(2k-1)!!=\frac{(2k)!}{k!2^k}=O\left(\sqrt{(2k)!}\right)\]
monomial terms in $f(x)$.

Here is the set of polynomials for the six-bit normalized standard
signature.  Note that if we fix the first bit as zero, we produce the
general form for all the five-bit normalized standard signatures, and
so forth.

\begin{eqnarray*}
f(000000) &=&\hspace{.11in}1 
\\
f(000011) &=&\hspace{.11in}\lambda_{2,1} 
\\
f(000101) &=&\hspace{.11in}\lambda_{3,1} 
\\
f(000110) &=&\hspace{.11in}\lambda_{3,2} 
\\
f(001001) &=&\hspace{.11in}\lambda_{4,1} 
\\
f(001010) &=&\hspace{.11in}\lambda_{4,2} 
\\
f(001100) &=&\hspace{.11in}\lambda_{4,3} 
\\
f(001111) &=&\hspace{.11in}\lambda_{4,1} \lambda_{3,2}  -\lambda_{4,2} \lambda_{3,1}  +\lambda_{4,3} \lambda_{2,1} 
\\
f(010001) &=&\hspace{.11in}\lambda_{5,1} 
\\
f(010010) &=&\hspace{.11in}\lambda_{5,2} 
\\
f(010100) &=&\hspace{.11in}\lambda_{5,3} 
\\
f(010111) &=&\hspace{.11in}\lambda_{5,1} \lambda_{3,2}  -\lambda_{5,2} \lambda_{3,1}  +\lambda_{5,3} \lambda_{2,1} 
\\
f(011000) &=&\hspace{.11in}\lambda_{5,4} 
\\
f(011011) &=&\hspace{.11in}\lambda_{5,1} \lambda_{4,2}  -\lambda_{5,2} \lambda_{4,1}  +\lambda_{5,4} \lambda_{2,1} 
\\
f(011101) &=&\hspace{.11in}\lambda_{5,1} \lambda_{4,3}  -\lambda_{5,3} \lambda_{4,1}  +\lambda_{5,4} \lambda_{3,1} 
\\
f(011110) &=&\hspace{.11in}\lambda_{5,2} \lambda_{4,3}  -\lambda_{5,3} \lambda_{4,2}  +\lambda_{5,4} \lambda_{3,2} 
\\
f(100001) &=&\hspace{.11in}\lambda_{6,1} 
\\
f(100010) &=&\hspace{.11in}\lambda_{6,2} 
\\
f(100100) &=&\hspace{.11in}\lambda_{6,3} 
\\
f(100111) &=&\hspace{.11in}\lambda_{6,1} \lambda_{3,2}  -\lambda_{6,2} \lambda_{3,1}  +\lambda_{6,3} \lambda_{2,1} 
\\
f(101000) &=&\hspace{.11in}\lambda_{6,4} 
\\
f(101011) &=&\hspace{.11in}\lambda_{6,1} \lambda_{4,2}  -\lambda_{6,2} \lambda_{4,1}  +\lambda_{6,4} \lambda_{2,1} 
\\
f(101101) &=&\hspace{.11in}\lambda_{6,1} \lambda_{4,3}  -\lambda_{6,3} \lambda_{4,1}  +\lambda_{6,4} \lambda_{3,1} 
\\
f(101110) &=&\hspace{.11in}\lambda_{6,2} \lambda_{4,3}  -\lambda_{6,3} \lambda_{4,2}  +\lambda_{6,4} \lambda_{3,2} 
\\
f(110000) &=&\hspace{.11in}\lambda_{6,5} 
\\
f(110011) &=&\hspace{.11in}\lambda_{6,1} \lambda_{5,2}  -\lambda_{6,2} \lambda_{5,1}  +\lambda_{6,5} \lambda_{2,1} 
\\
f(110101) &=&\hspace{.11in}\lambda_{6,1} \lambda_{5,3}  -\lambda_{6,3} \lambda_{5,1}  +\lambda_{6,5} \lambda_{3,1} 
\\
f(110110) &=&\hspace{.11in}\lambda_{6,2} \lambda_{5,3}  -\lambda_{6,3} \lambda_{5,2}  +\lambda_{6,5} \lambda_{3,2} 
\\
f(111001) &=&\hspace{.11in}\lambda_{6,1} \lambda_{5,4}  -\lambda_{6,4} \lambda_{5,1}  +\lambda_{6,5} \lambda_{4,1} 
\\
f(111010) &=&\hspace{.11in}\lambda_{6,2} \lambda_{5,4}  -\lambda_{6,4} \lambda_{5,2}  +\lambda_{6,5} \lambda_{4,2} 
\\
f(111100) &=&\hspace{.11in}\lambda_{6,3} \lambda_{5,4}  -\lambda_{6,4} \lambda_{5,3}  +\lambda_{6,5} \lambda_{4,3} 
\\
f(111111) &=&\hspace{.11in}\lambda_{6,1} \lambda_{5,2} \lambda_{4,3}  -\lambda_{6,1} \lambda_{5,3} \lambda_{4,2}  +\lambda_{6,1} \lambda_{5,4} \lambda_{3,2} \\
 &&  -\lambda_{6,2} \lambda_{5,1} \lambda_{4,3}  +\lambda_{6,2} \lambda_{5,3} \lambda_{4,1}  -\lambda_{6,2} \lambda_{5,4} \lambda_{3,1} \\
 &&  +\lambda_{6,3} \lambda_{5,1} \lambda_{4,2}  -\lambda_{6,3} \lambda_{5,2} \lambda_{4,1}  +\lambda_{6,3} \lambda_{5,4} \lambda_{2,1} \\
 &&  -\lambda_{6,4} \lambda_{5,1} \lambda_{3,2}  +\lambda_{6,4} \lambda_{5,2} \lambda_{3,1}  -\lambda_{6,4} \lambda_{5,3} \lambda_{2,1} \\
 &&  +\lambda_{6,5} \lambda_{4,1} \lambda_{3,2}  -\lambda_{6,5} \lambda_{4,2} \lambda_{3,1}  +\lambda_{6,5} \lambda_{4,3} \lambda_{2,1} 
\end{eqnarray*}

\section{Matchgate Recursion} \label{app:matchgate}
The body of this paper has focussed on an algebraic recursion that
allowed us to construct $(n+1)$-bit standard signatures out of $n$-bit
standard signatures.  One may wonder if there is a planar matchgate
counterpart-- that is, is there some sort of recursive planar matchgate
structure that reflects this.  There is, and we offer one such
possibility below.  We begin by reviewing a particularly useful
4-bit standard signature.

\begin{lemma} \label{lem:switch}
Define the switch function $f_{switch}:V^4\rightarrow \FF$
as:
\[f_{switch}(0000)=f_{switch}(0101)=f_{switch}(1010)=1\]
\[f_{switch}(1111)=-1\]
Then the switch function is a standard signature.
\end{lemma}

\proof It is possible to prove this result only using algebra, but it
is simpler to construct the switch matchgate directly.  These planar
matchgates are modified versions of Figure 8 from
Valiant~\cite{valiant3}.

First, suppose that $\FF$ is not characteristic two.  Then
$\frac{1}{2}\in \FF$, and we can consider the following planar matchgate
(where unmarked edges have weight 1):

\begin{center}
\epsfig{file=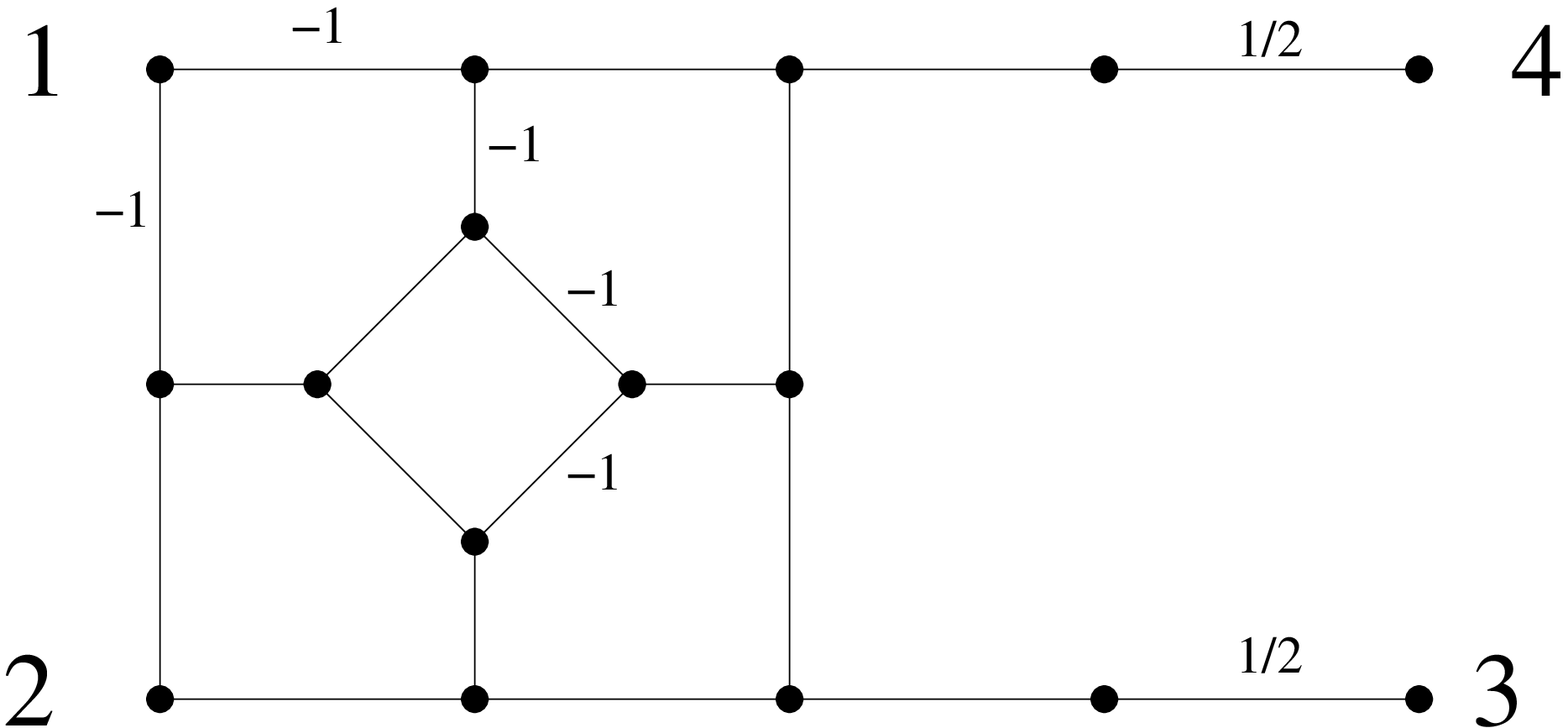, height=1.5in}
\end{center}

On the other hand, if $\FF$ has characteristic two, we can use the
following planar matchgate (where all edges have weight 1):

\begin{center}
\epsfig{file=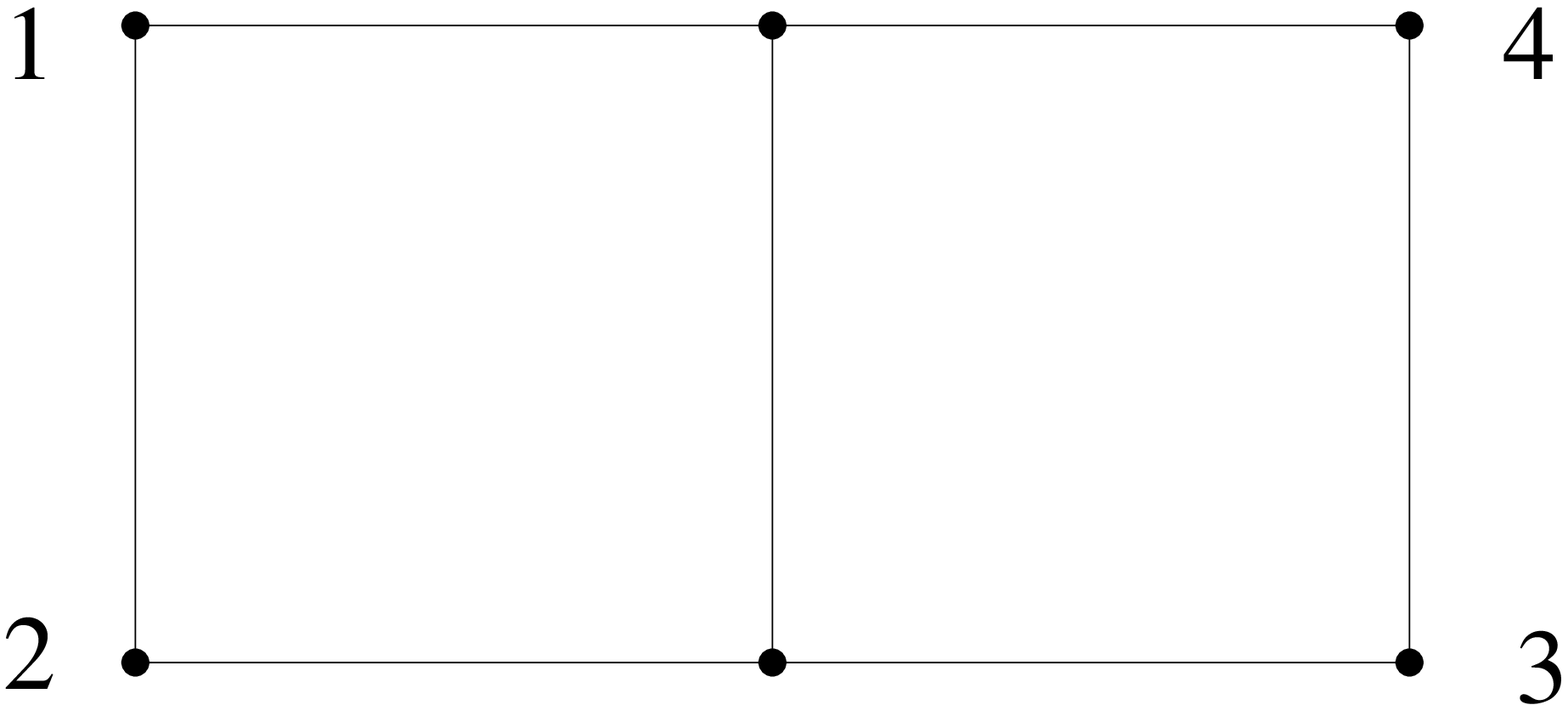, height=1.5in}
\end{center}

If we count up the weighted perfect matchings for these two graphs
over their respective fields, they produce the switch function, as
desired.\eop

To simplify our diagrams, we will (following Valiant) adopt the
following emblem for the switch matchgate, where the underlying
planar matchgate is chosen from the two above depending on the base field's
characteristic:
\begin{center} \label{diagram:switch}
\epsfig{file=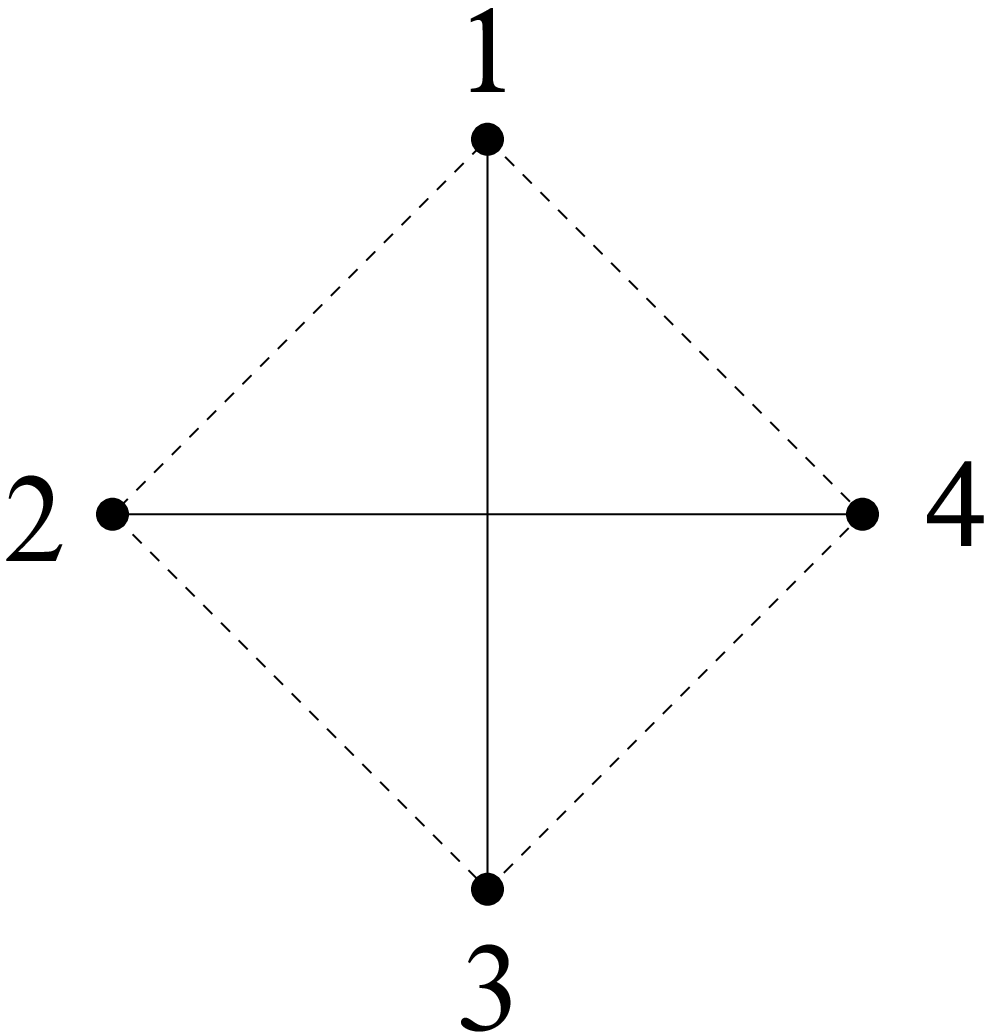, height=1.5in}
\end{center}
Since the outputs are symmetric under rotation in the plane, we can
ignore the labels without causing any ambiguity.  If we consider
input/output nodes 1 and 3, observe that either they must both be
saturated, or neither of them is saturated.  In other words, the
planar matchgate acts as though there were a ``virtual edge'' between the
nodes.  The same principle applies to nodes 2 and 4.

Notice that over fields of characteristic two, the switch matchgate is
equivalent to letting two edges cross each other (since
$f_{switch}(1111)=-1\equiv 1 \bmod 2$).  In other words, for those
particular fields, the planarity requirement in a planar matchgate is
redundant; we can simply take any non-planar crossings and replace them
with planar switch matchgates.  We can state this corollary formally:

\begin{corollary}
If we operate over a field $\FF$ of characteristic 2, we can remove
the planarity restriction from the definition of a planar matchgate without
changing the resulting set of standard signatures.
\end{corollary}

In any event, the purpose of this section is to provide a recursive
planar matchgate construction that mirrored the algebraic recursion
from Subsection~\ref{subsec:recursion}.  Here is one example, where we
choose arbitary $\lambda_i\in\FF$, and all unmarked edges have weight
one.

\hspace{-.5in}\epsfig{file=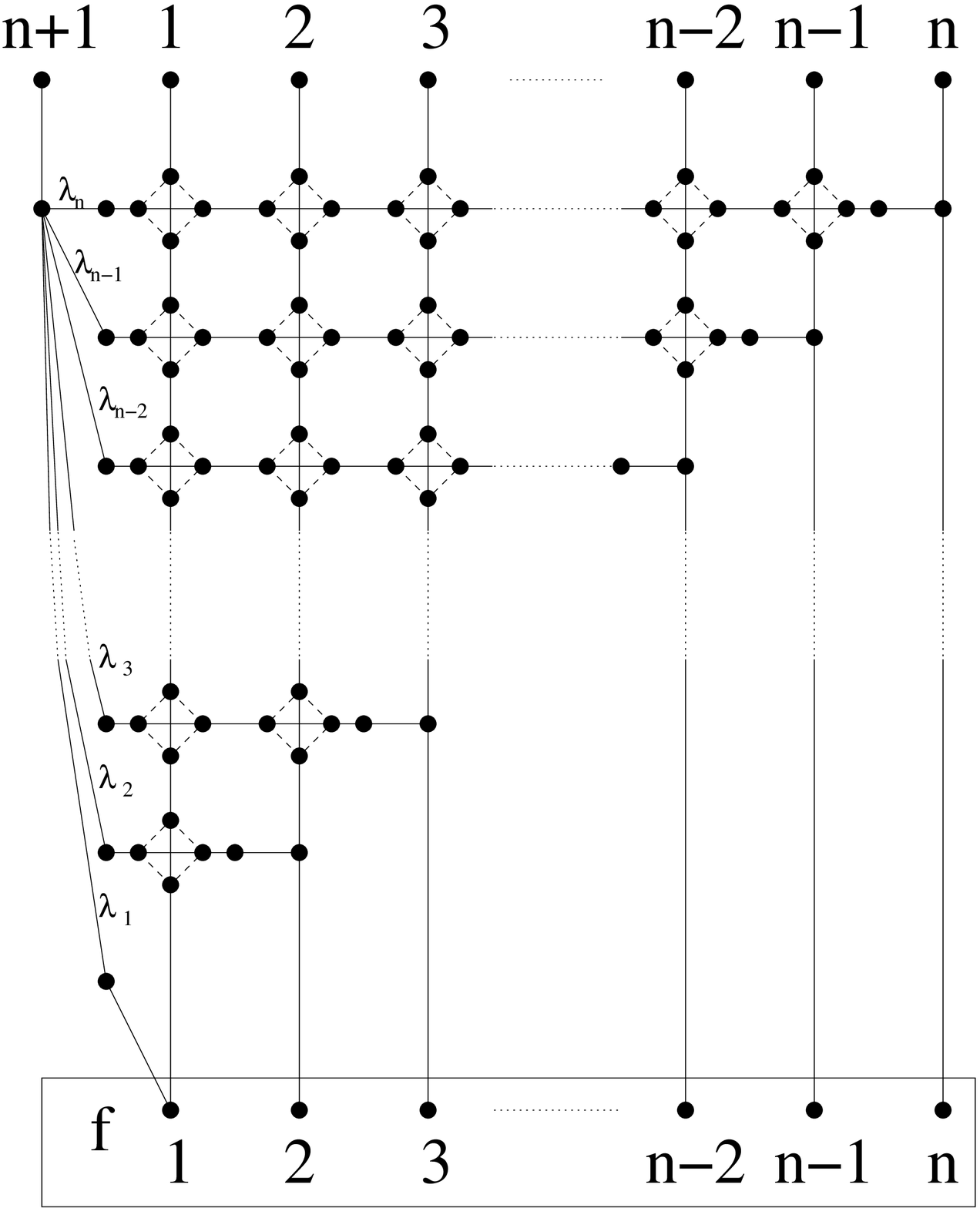, height=6in}

\bibliography{}

\end{document}